# LACK – a VoIP Steganographic Method

Wojciech Mazurczyk and Józef Lubacz
Warsaw University of Technology, Institute of Telecommunications
Warsaw, Poland, 00-665, Nowowiejska 15/19

**Abstract.** The paper presents a new steganographic method called LACK (**L**ost **A**udio Pa**CK**ets Steganography) which is intended mainly for VoIP. The method is presented in a broader context of network steganography and of VoIP steganography in particular. The analytical results presented in the paper concern the influence of LACK's hidden data insertion procedure on the method's impact on quality of voice transmission and its resistance to steganalysis.

## 1. Introduction

Communication network steganography is a method of hiding secret data in users' normal data transmissions, ideally, so it cannot be detected by third parties. One of the most popular steganographic techniques is to use a covert channel, which enables manipulating certain properties of the communications medium in an unexpected, unconventional, or unforeseen way. In the past few years the interest in steganographic methods that may be used in computer networks has grown considerable, mostly due to presumed usage of hidden communication by terrorists [27]. Many new methods have been proposed and analyzed, e.g. [1], [2] or [3]. In this paper we propose a classification of network steganography methods. On this background we present and analyse the performance of a new steganographic method called LACK (**L**ost **A**udio Pa**CK**ets Steganography) which was recently proposed and filed for patenting [23]. The general idea of the method was described in [24]. LACK is intended for a broad class of multimedia, real-time applications, but its main foreseen application, at least for now, is IP telephony. The method utilizes the fact that for usual multimedia communication protocols like RTP (Real-Time Transport Protocol) excessively delayed packets are not used for reconstruction of transmitted data at the receiver, i.e. the packets are considered useless and discarded.

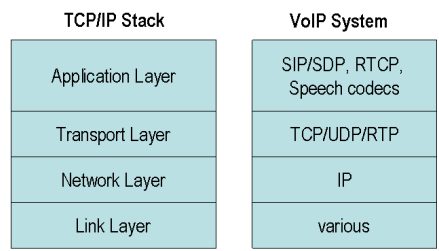

Fig. 1 VoIP stack and protocols

Voice over IP (VoIP), or IP telephony, is one of the services of the IP world which is changing the entire telecommunications landscape. Because of its popularity, it is becoming a natural target for steganography. We propose to name steganographic techniques applied to VoIP traffic *steganophony*. This term, if accepted, would pertain to information hiding techniques in any layer of the TCP/IP protocol-stack (Fig. 1), including methods such as audio watermarking and techniques applied in speech codecs.

For VoIP systems, four possible hidden communication scenarios may be considered, as illustrated in Fig. 2. The first scenario (1 in Fig. 2) is most common: the sender and the receiver perform VoIP conversation while simultaneously exchanging steganograms. The conversation path is the same as the hidden data path. In the next three scenarios (marked 2-4 in Fig. 2) only a part of the VoIP end-to-end path is used for hidden communication as a result of actions undertaken by intermediate nodes; the sender and/or receiver are, in principle, unaware of the steganographic data exchange.

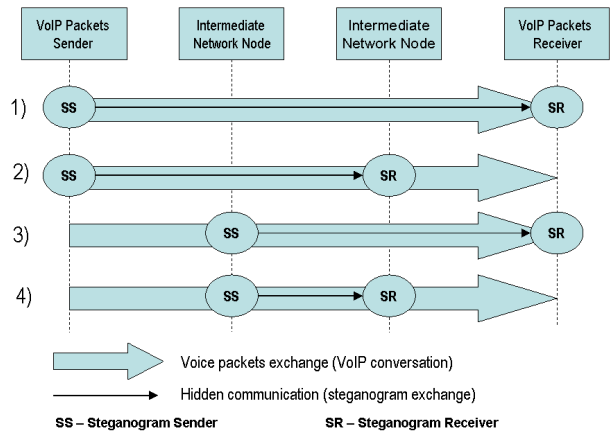

Fig. 2 Hidden communication scenarios for VoIP

Steganophony may be classified into three groups (Fig. 3):

**(S1) Steganographic methods which modify packets** – network protocol headers or payload fields. Examples of such solutions include (a) modifications of free/redundant headers' fields of IP, UDP or RTP [10] protocols during conversation phase and (b) modification of signaling messages in e.g. SIP [11]. Information hiding which is based on affecting packets' payload usually uses digital audio watermarking algorithms, e.g. DSSS [12] or QIM [13].

**(S2) Steganographic methods which modify packets' time relations**, e.g. by affecting the sequence order of RTP packets, modifying their inter-packet delay or by introducing intentional losses.

**(S3) Hybrid steganographic methods** which modify both the content of packets and their time relations. An example of such solution is the LACK method, which is presented in this paper.

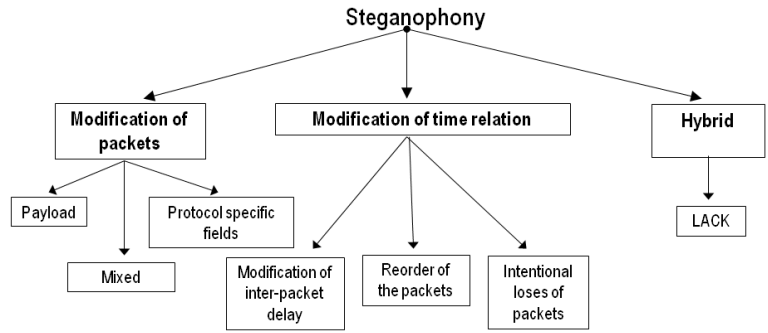

Fig. 3 Steganophony classification

Examples of methods from group S1:
- Methods which modify protocols specific fields – SIP, SDP [16], RTP, RTCP [10] (VoIP specific protocols) and additionally: IP, TCP, UDP (network specific protocols).
- Methods which modify packets' payload: audio watermarking algorithms, speech codec steganographic techniques (e.g. using SID frames or codec specific steganographic methods).
- Mixed techniques: HICCUPS (<u>Hi</u>dden <u>C</u>ommunication System for <u>C</u>orr<u>u</u>pted Network<u>s</u>, [4]).

Some characteristic features of methods from group S1:
- Steganographic methods which use protocol specific fields usually yield relatively high steganographic capacity. Implementation and detection is relatively straightforward. Drawback: potential loss of some of the protocols' functionality.

- Steganographic methods which utilize payload of the packets generally yield lower steganographic capacity and are harder to implement and detect. Drawback: potential deterioration of voice quality.
- Mixed techniques offer high steganographic capacity, but the implementation is harder due to required low-level access to hardware. For the same reason steganalysis is harder to perform. Drawback: increased frame error rate.

Examples of the steganographic methods from group S2:
- Methods which affect sequence order of packets [15] (in VoIP possible only for RTP).
- Methods which modify inter-packet delay [14] (in VoIP possible for RTP and RTCP; for some protocols, e.g. SIP, not useful due to small number of messages).
- Methods which introduce intentional losses by skipping sequence numbers at the sender [5] (for RTP and RTCP protocols).

Some characteristic features of methods from group S2:
- Sender-receiver synchronization required.
- Lower steganographic capacity and harder to detect than in the case of methods which utilize protocol specific fields.
- Straightforward implementation.
- Drawback: potential deterioration of conversation quality.

As already mentioned, LACK is an example of hybrid steganographic methods from group S3 which modify both packets and their time dependencies. In section 2 LACK is presented in some detail and performance issues involved in using the method are discussed. Section 3 introduces and analyses a procedure for inserting hidden data; the procedure takes account of the estimated voice call duration. In section 4 the influence of using LACK on voice QoS is analysed. Section 5 concludes our work and indicates possible future research.

## 2. LACK – The Idea and Performance Issues

At the transmitter, some selected audio packets are intentionally delayed before transmitting. If the delay of such packets at the receiver is considered excessive, the packets are discarded by a receiver which is not aware of the steganographic procedure. The payload of the intentionally delayed packets is used to transmit secret information to receivers aware of the procedure, so no extra packets are generated. For unaware receivers the hidden data is "invisible".

LACK may be used in four basic scenarios illustrated in Fig. 4. In scenario (1), one packet is selected from the RTP stream and its voice payload is substituted with bits of the steganogram. In scenario (2) chosen packets are delayed by a certain time and then sent through the communication channel. In scenario (3), if an excessively delayed packet reaches a receiver unaware of the steganographic procedure, it is discarded. In scenario (4), if the receiver knows about the hidden communication, then instead of deleting the packet the receiver extracts the payload.

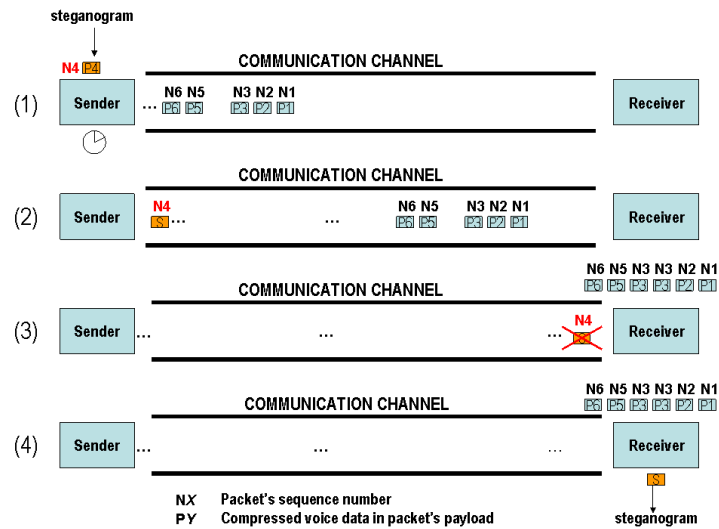

Fig. 4. LACK scenarios

LACK, although it is an application layer steganography technique, is less complex to implement than most audio steganography algorithms. This is due to the fact that RTP is usually integrated in telephone endpoints (softphones) so access to RTP packets generation and modification is easier to perform than in the case of lower layer protocols like IP or UDP.

The performance of LACK depends on many factors such as the details of the communication procedure (in particular the type of codec used, the size of the voice frame, the size of the receiving buffer, etc.) and on the network QoS (packet delay, packet loss probability and jitter). We discuss these issues in the following.

No real-world steganographic method is perfect – whatever the method, the hidden information can be potentially discovered. In general, the more hidden information is inserted into the data stream, the greater the chance that it will be detected, e.g. by scanning the data flow or by some other steganalysis methods. Moreover, the more audio packets are used to send covert data, the greater the potential deterioration of the quality of VoIP connection. Thus the procedure of inserting hidden data has to be carefully chosen and controlled in order to minimize the chance of detecting inserted data and to avoid excessive deterioration of QoS.

To avoid excessive deterioration of QoS lost packet ratio must be kept below a certain, accepted level. This level depends on the speech codec. For example, according to [20], maximum loss tolerance is 1% for G.723.1, 2% for G.729A and 3% for G.711 codecs. If a special mechanism to deal with lost packets at the receiver is utilized, e.g. the PLC (Packet Loss Concealment) [21], then the acceptable level of lost packets e.g. for G.711 codecs increases from 3% to 5%. Thus the amount of steganographic data that can be inserted by LACK, and in effect the additional packet loss introduced by LACK, depends on the acceptable level of the total packet loss. For example, for the G.711 speech codec with data rate 64 kbit/s and data frame size of 20 ms, if the packet loss probability introduced by the LACK procedure is 0.5%, then the theoretical hidden communication rate is 320 b/s.

To guarantee that the an audio packet will be recognized as lost by receiver, it must be excessively delayed by the LACK procedure. To set this delay, the size of the receiver's de-jitter buffer must be taken into account. A de-jitter buffer is used to alleviate the jitter effect, i.e. the variations in packets arrival time caused by queuing, contention and serialization in the network. The size of the buffer is implementation-dependent. It may be fixed or adaptive, and is usually between 30 and 70 ms; RTP packet will be recognized as lost when the delay is greater than the delay introduced by the de-jitter buffer. LACK users have to exchange information about the sizes of their de-jitter buffers before starting the steganographic procedure. To limit the risk of detection of the hidden data, the delay chosen by LACK users should be as low as possible.

Additionally, packets losses introduced by the network must be carefully monitored. Because LACK uses legitimate RTP traffic, thus it increases overall packets losses. To ensure that the total packet loss introduced by the network and by LACK will not degrade perceived quality of the conversation, the level of the lost packets used for steganographic purposes must be controlled and dynamically adapted.

Steganalysis of LACK is harder to perform than for most other steganographic methods that were described in Section 1. Packet loss in IP networks is a "natural phenomenon", so intentional losses introduced by LACK are not easy to detect, if kept on a reasonable level. Potential LACK steganalysis methods include:

- Statistical analysis of lost packets for calls in some sub-network. This type of steganalysis may be implemented with a passive warden [22] (or some other network node), based e.g. on information included in RTCP reports (cumulative number of packets lost field) exchanged between users during their communication or by observing RTP streams flow (packets' sequence numbers). If for some of the observed calls the number of lost packets is higher than average (or some chosen threshold) this may be used as an indication of potential use of LACK.

- Statistical analysis based on VoIP calls duration. If the call duration probability distribution for a certain sub-network is known, then statistical steganalysis may be performed to discover VoIP sources that do not fit to the distribution (the duration of LACK calls may be longer than non-LACK calls in effect of introducing steganographic data).

- An active warden [22] which analyses all RTP streams in the network (*SSRC* identifier and fields: *Sequence Number* and *Timestamp* from RTP header) can identify packets that are already too late to be used for voice reconstruction. The active warden may erase their payloads fields or simply drop them. A potential problem which arises in this case is to avoid eliminating delayed packets that still may be used for conversation reconstruction. The size of the jitter buffer at the receiver is, in principle, unknown to the active warden. If an active warden drops all delayed packets, then it will potentially drop packets that still can be useful for voice reconstruction. In effect, the quality of conversation may deteriorate considerably. Moreover, not only steganographic calls are affected but also non-steganographic ones are "punished".

What follows from the performance issues mentioned above is that the LACK procedure of inserting hidden data has to be carefully designed. The focal point of this paper is the design and analysis of hidden data insertion rate *IR* [bits/s]. *IR* depends on the amount of hidden data to be sent and on the call duration. In principle, the call duration may be adjusted to the amount of hidden data to be sent. This, however, could cause that the distribution of calls applying LACK differs from the call duration distribution of LACK-less calls, and as a consequence make LACK vulnerable to statistical steganalysis based on call duration. Thus rather than adjusting the call duration to the amount of hidden data to be sent, it is preferable to adjust the hidden data insertion rate *IR* to LACK-less calls duration probability distribution. This, in turn, requires making *IR* dependent on that distribution. Obviously, this is especially important in the case of a predefined group of frequent LACK users, rather than in the case of sporadic use of LACK. In the present paper we focus on the former case.

It should be emphasised that the hidden data insertion procedures introduced and analysed in this paper can be utilized by decent LACK users who use their own VoIP calls to exchange covert data, but also by intruders who are able to covertly send data using third party VoIP calls (e.g. in effect of earlier successful attacks by using trojans or worms or by distributing modified versions of a popular VoIP software). This is a usual tradeoff requiring consideration in a broader steganography context which is beyond the scope of this paper.

## 3. Hidden Data Insertion Rate (IR)

In the following analysis we consider the dependence of the hidden data insertion rate *IR* for a particular call on the elapsed time of that call, i.e. we consider *IR* that is made time dependent. As shown in our analysis, such time-dependent *IR* procedure allows for decreasing the *IR* during the call duration, compared to the *IR* at call initiation time. In effect, the negative influence of LACK on QoS can be decreased and resistance to steganalysis increased, especially for call duration distributions with coefficient of variation much greater than 1. Available experimental data concerning VoIP call duration distributions seem to indicate that this is realistic for real-life VoIP calls. Our goal in this section is to express *IR* with the coefficient of variation for possibly wide range of call duration distributions.

For PSTN the call duration probability distribution was well known due to extensive experimental research. For many decades the exponential distribution was assumed a good enough approximation for engineering purposes. VoIP is a relatively new service and thus only few reliable experimental data is available, so in many research papers concerning IP voice traffic (e.g. [7], [8], [9]) the exponential call duration is still assumed. Current experiments prove however that this assumption is far from being realistic.

Birke et al. [6] captured real VoIP traffic traces (about 150 000 calls) from FastWeb, an Italian telecom operator. The obtained call duration probability distribution is reproduced in Fig. 5 with a solid line. To illustrate qualitatively the degree in which the experimental results differ from exponential distribution and some other chosen distributions (hyperexponential and log-normal), these are drawn with broken lines in Fig. 5. As can be seen, the differences are considerable and no straightforward approximation of the experimental data with standard distributions is available.

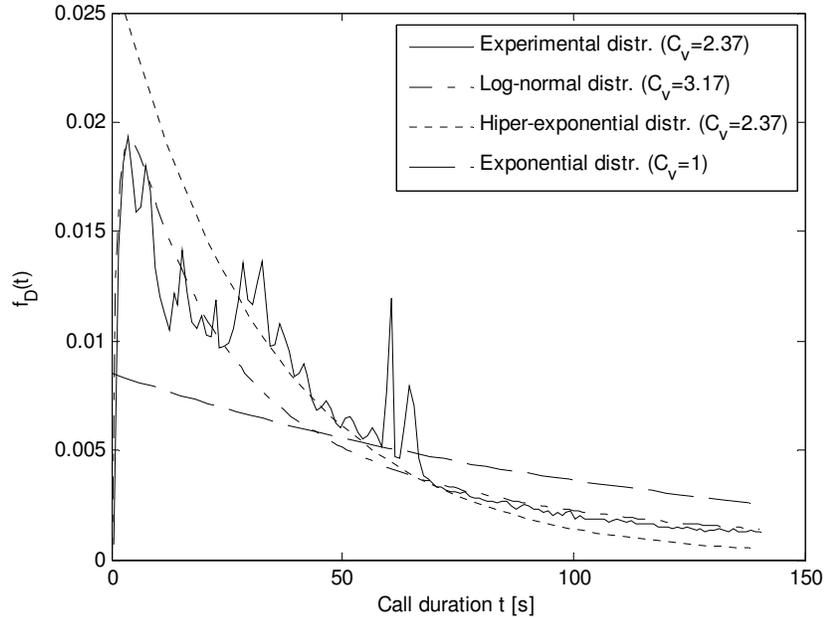

Fig. 5. VoIP call duration – comparison of experimental distribution with selected standard probability distributions

The experimental data from [6] yields average call duration $E(D) = 117.31$ and standard deviation $\sigma(D) = 278.74$, thus the coefficient of variation $C_V = \sigma(D)/E(D) = 2.37$ (for the exponential distribution $C_V= 1$).

To achieve an analytic approximation of the experimental data a combination of some standard distributions can be used, for example:

$$f_D(t) = \begin{cases} \dfrac{1}{1.55t\sqrt{2\pi}} e^{-\dfrac{(\ln(t)-3.8)^2}{4.805}} & for \quad 0 \leq t < 27.5 \\ 0.000114 e^{-0.00114\,t} + 0.027252 e^{-0.03028\,t} & for \quad 66.5 < t \leq 27.5 \\ \dfrac{1}{1.55t\sqrt{2\pi}} e^{-\dfrac{(\ln(t)-3.8)^2}{4.805}} & for \quad 66.5 \leq t \leq 455 \end{cases} \quad (1)$$

The above analytic approximation is quite complex and of little practical use for our purposes, i.e. for establishing the dependence of the insertion rate *IR* on some simple enough characterization of the call duration distribution.

A reasonably wide range of call distribution types can however be achieved and effectively analysed/used with the 2-parameter Weibull distribution and appropriately chosen parameters: the shape parameter $k > 0$ and the scale parameter $\lambda > 0$. The complementary cumulative probability distribution function ($\overline{F}_D$) and probability density function ($f_D$) are as follows:

$$\overline{F}_D(t;k,\lambda) = e^{-\left(\dfrac{t}{\lambda}\right)^k}$$

$$f_D(t;k,\lambda) = \dfrac{k}{\lambda}\left(\dfrac{t}{\lambda}\right)^{k-1} e^{-\left(\dfrac{t}{\lambda}\right)^k} \quad (2)$$

The $\lambda$ parameter was set so to achieve the above experimental average call duration time $E(D) = 117.31$ and the $k$ parameter was varied so to obtain a wide range of $C_V$ values. In Tab. 1 the analyzed values are summarized.

| Weibull parameters | k=3.4, λ=130.57 | k=2, λ=132.37 | k=1.2, λ=124.71 | k=1, λ=117.31 | k=0.8, λ=103.54 | k=0.6, λ=77.97 | k=0.5, λ=58.65 | k=0.4, λ=35.3 |
|---|---|---|---|---|---|---|---|---|
| $C_V$ | 0.32 | 0.52 | 0.84 | 1 | 1.26 | 1.76 | 2.23 | 3.14 |

Table 1. Weibull distribution parameters $k$ and $\lambda$ and corresponding $C_V$ values.

In Fig. 6 the Weibull probability distribution is depicted for the parameters from Tab. 1 to illustrate the resulting wide range of distribution shapes. Note by the way that for $k = 1$ the Weibull distribution equals the exponential distribution ($C_V = 1$), for $k = 2$ it becomes the Rayleigh distribution ($C_V = 0.52$) and for $k = 3.4$ it resembles the normal distribution ($C_V = 0.32$).

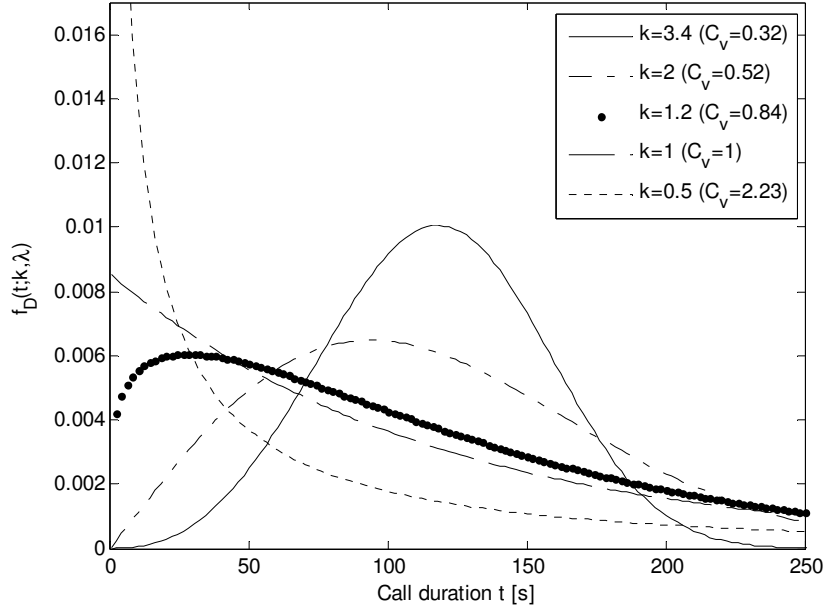

Fig. 6. Weibull distribution for various $k$, $\lambda$ and $C_V$

For an arbitrary instant of a call the average residual call duration is well know to be equal

$$E(R) = \frac{E(D^2)}{2E(D)} \quad (3)$$

or equivalently

$$E(R) = \frac{C_V^2 + 1}{2} E(D) \quad (4)$$

Suppose that at the beginning of a call the insertion rate is set to $IR = S/E(D)$, where $S$ is amount of data to be sent covertly. If $C_V > 1$ then $E(R) > E(D)$, which seems to be the case for VoIP real-world calls as indicated above, then beginning from some arbitrary instant of the call we may decrease the insertion rate to $IR = S/E(R)$, which is beneficial from the point of view of call quality and resistance to detection of the hidden data.

The above indicates that it is reasonable to make the insertion rate dependent on the elapsed time of a call. It is nevertheless not practical to use the classical definition of residual call duration since it involves an arbitrary time instant and not the current call duration. We are rather interested in the expected call duration on condition it has already lasted $t$ units of time:

$$E(D\,|\,D > t) = \frac{1}{P(D > t)} \int_t^\infty x f_D(x) dx = t + \frac{1}{\overline{F}_D(t)} \int_t^\infty \overline{F}_D(x) dx \quad (5)$$

and

$$t \leq E(D\,|\,D > t) \leq \frac{E(D)}{\overline{F}_D(t)} \quad (6)$$

For the Weibull distributions considered above

$$E(D \mid D > t) = t + e^{\left(\frac{t}{\lambda}\right)^k} \int_t^\infty e^{-\left(\frac{x}{\lambda}\right)^k} dx \qquad (7)$$

and

$$t \leq E(D \mid D > t) \leq e^{\left(\frac{x}{\lambda}\right)^k} \lambda \Gamma\left(1 + \frac{1}{k}\right) \qquad (8)$$

For chosen parameters from Tab.1 we obtain results shown in Fig. 7. The figure shows also the $E(D|D>t)$ function for the experimental data presented in Fig. 5.

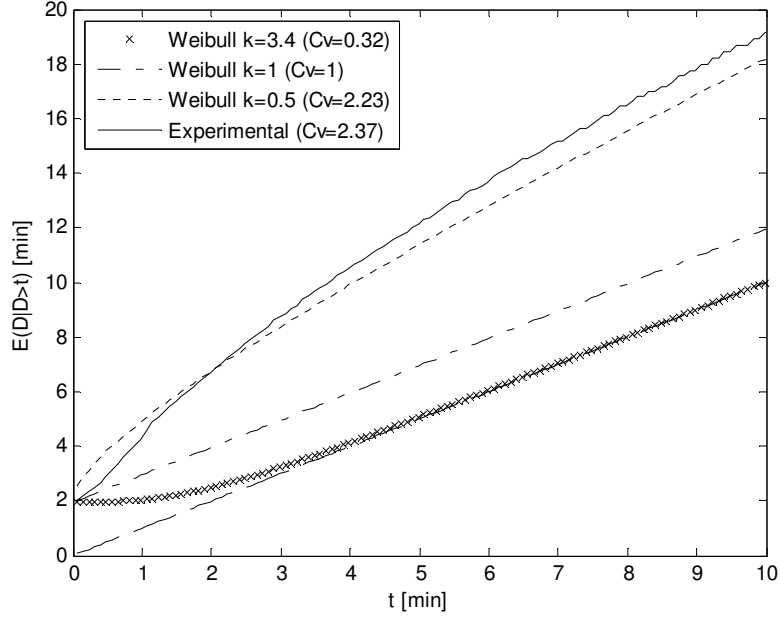

Fig. 7. $E(D|D>t)$ for different Weibull distributions and for the experimental data distribution

The curves from Fig. 7 may be approximated with good accuracy as follows:

$$E(D \mid D > t) \approx 1.32 C_v + t\sqrt{C_v} + 0.59 \quad [\text{min}] \qquad (9)$$

If $S_R(t)$ is the amount of data remaining to be sent covertly at instant $t$ of the call, then the insertion rate at time $t$ is:

$$IR(t) = \frac{S_R(t)}{E(D \mid D > t)} \qquad (10)$$

$$S_R(t) = S - \int_0^t IR(x) dx$$

Based on results presented in Fig. 7 and eq. 10, assuming $S = 1000$ bits, the $IR(t)$ functions are as presented in Fig. 8.

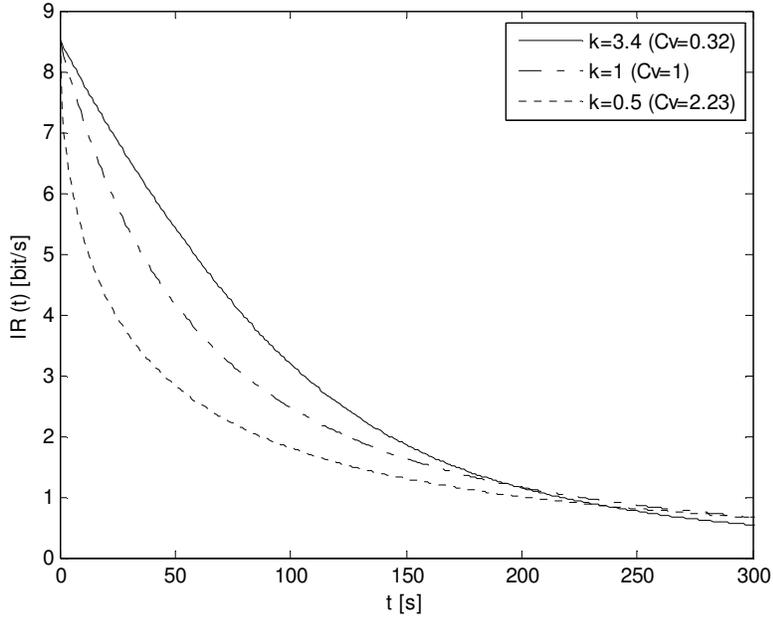

Fig. 8. *IR(t)* for chosen Weibull distributions, *S=1000* bits

As can be seen from Fig. 8, the hidden data insertion rate *IR(t)* may be decreased during the call duration compared to its initial value *S/E(D)*. This is the desired effect which was aimed at: as the call proceeds, the *IR* is adjusted – decreased – according to the expected remaining duration of the call, which is, as already mentioned, beneficial from the point of view of voice quality and resistance to steganalysis. This feature is illustrated qualitatively in Fig. 9; in quantitative terms the feature is expressed by eq. 11.

$$X(t) = IR(0) - IR(t) = \frac{S}{E(D)} - \frac{S - \int_0^t IR(x)dx}{E(D\,|\,D>t)} \qquad (11)$$

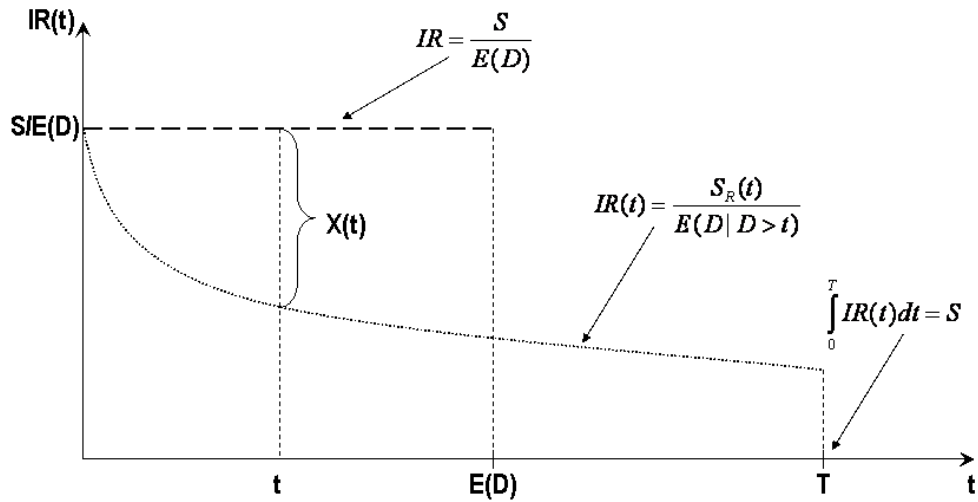

Fig. 9. The effect of using *IR(t)* based on *E(D|D>t)*

## 4. Dependence of the Insertion Rate IR on Estimated Call Quality

In this section we focus on the dependence of the insertion rate *IR* on estimated call quality resulting from packet loss. Call quality may be expressed in terms of subjective and objective quality measures. Objective measures are usually based on algorithms such as the E-Model [18], PAMS or PESQ [26]. The objective measures can be transformed into subjective quality measures. In our analysis we shall use the subjective measure MOS (Mean Opinion Score) [25] which according to [19] can be related to packet loss probability $p_{loss}$ as follows:

$$MOS = \alpha \cdot \exp(\beta \cdot p_{loss}) + \gamma \qquad (12)$$

where $\alpha$, $\beta$ and $\gamma$ are network/service-type dependent parameters; for Skype telephony the parameters were evaluated to be [19]: $\alpha = 3.0829$, $\beta = -4.6446$ and $\gamma = 1.07$.

Since LACK introduces additional packet loss $p_{LACK}$ then in the above equation $p_{loss}$ should be substituted with $p_{loss} + p_{LACK}$:

$$MOS = \alpha \cdot \exp(\beta \cdot (p_{loss} + p_{LACK})) + \gamma \qquad (13)$$

Fig. 10 shows the dependence of MOS on $p_{loss}$ for different values of $p_{LACK}$, assuming $\alpha$, $\beta$ and $\gamma$ values estimated for Skype telephony.

Taking into account the call-duration insertion rate *IR(t)* principle introduced in the previous section and that

$$p_{LACK}(t) = \frac{IR(t)}{N_P \cdot P_P} \qquad (14)$$

where $N_P$ is the number of RTP packets generated in a unit of time and $P_P$ is the length of a RTP packet data field, then

$$MOS(t) = \alpha \cdot \exp\left(\beta \cdot \left(p_{loss} + \frac{IR(t)}{N_P \cdot P_P}\right)\right) + \gamma \qquad (15)$$

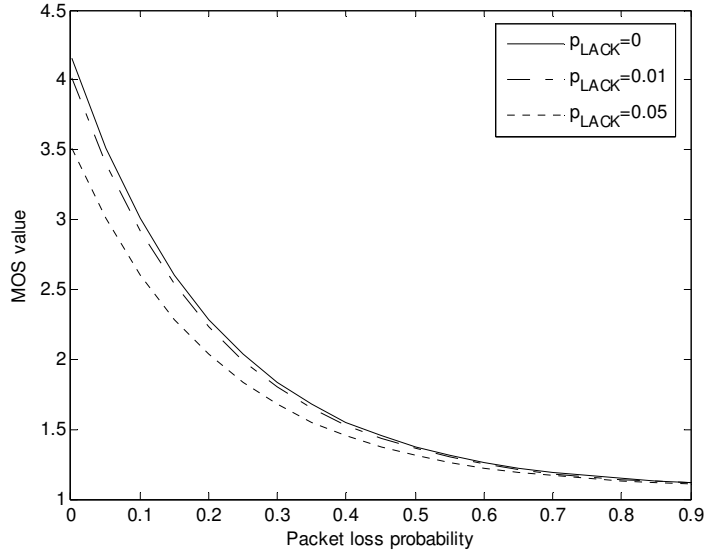

Fig. 10. MOS dependence on $p_{loss}$ and $p_{LACK}$ for Skype telephony

If the MOS probability distribution for a considered network in which LACK is to be used is known, then the above equation can be used to set appropriate *IR(t)* values so that the additional deterioration of voice quality introduced by LACK is kept below a desired threshold.

Fig. 11 presents the MOS probability distribution for a VoIP network based on experimental data from [6]. If the shape of the presented curve could be considered typical, then establishing an appropriate adjustment of *IR(t)* based on eq. 15 is rather straightforward.

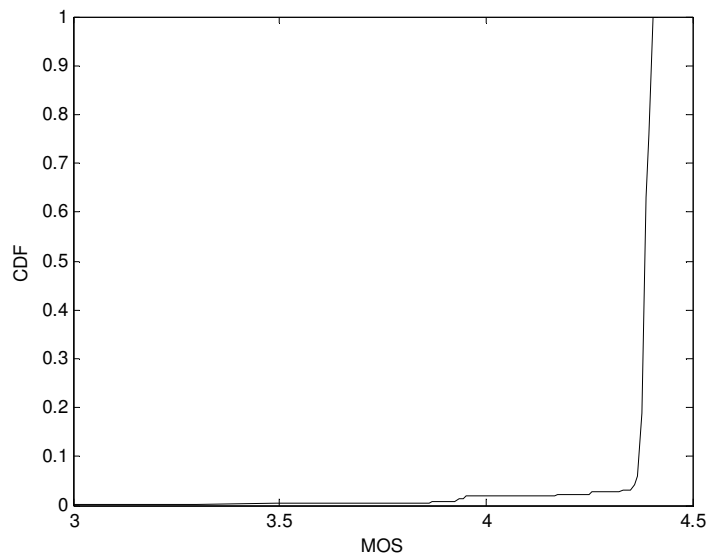

Fig. 11. MOS probability distribution (experimental data [6])

An alternative to the approach described above is to adjust *IR(t)* based on online measurement of voice quality during the call. Such an approach would require online exchange of information on voice quality parameters between the sender and the receiver, e.g. with the use of the RTCP protocol (Sender Reports and Receiver Reports [10] or Extended Reports [17]).

## 5. Conclusions and Future Work

The LACK steganographic method is a new idea which requires detailed performance evaluation. This paper is an initial step in this direction. We have focused mainly on the hidden data insertion rate *IR* procedure and its dependence on estimated average call duration and voice quality. It was shown that the insertion rate may be effectively made dependent on the current call duration time, and that this dependence can be expressed with good accuracy with the coefficient of variation of the call duration probability distribution. We have also derived analytical relations which enable making *IR(t)* dependent on voice quality parameters. All derived formulae are simple and can be straightforwardly implemented.

The effectiveness of the resulting hidden data insertion procedure will depend on the accuracy of the estimated mean call duration, the coefficient of variation of the call duration and the probability distribution of voice quality for the network (sub-network) which is intended to be used for sending steganographic data with the LACK method. Thus to evaluate realistically this effectiveness more experimental data has to gathered, nevertheless the authors believe that the analysis presented in this paper indicates that LACK provides good chance for high effectiveness.

Because LACK is a hybrid steganographic method in the sense of the classification introduced in this paper, then the steganographic bandwidth it can provide may be expected to be comparable or slightly lower than for methods which modify packets only, and higher than for methods which only modify packets' time relations. However to determine the effective, real-world steganographic bandwidth that LACK can provide, more research has to be undertaken. In particular, a more realistic, experimental based data concerning network QoS parameters have to be gathered.

Moreover, more sophisticated (than presented in this paper) hidden data insertion procedures could be considered. The idea of making the insertion rate time dependent can be further developed, i.e. by for example considering the probability *P(D>x|D>t)*, *x>t*, rather than *E(D|D>t)* as is the case in the analysis presented in this paper. Also the dependence of the hidden data insertion rate *IR* on call quality measures can be extended in order to take into account not only packet loss, but also packet delay and jitter. The alternative is to adjust *IR(t)* on the basis of online measurements of voice quality parameters during the call duration.

Analytical and experimental results concerning the above issues will be presented by the authors in forthcoming papers.

**ACKNOWLEDGMENTS**

- This research was partially supported by the Ministry of Science and Higher Education, Poland (grant no. 3968/B/T02/2008/34).
- The authors would like to thank R. Birke, M. Mellia, M. Petracca and D. Rossi from Politecnico di Torino for sharing details of their VoIP experimental data.